\edef\restoreinteractionmode{\interactionmode=\the\interactionmode\relax}
\documentclass[rapid]{jfm}
\restoreinteractionmode

\usepackage{amsmath}
\usepackage{amssymb}
\usepackage{accents}         
\usepackage{subcaption}      
\usepackage{comment}
\usepackage{tikz}
\usetikzlibrary{decorations.markings,calc,shapes.geometric}
\usepackage{pgfplots}
\pgfplotsset{compat=1.18}
\usepackage[bbgreekl]{mathbbol} 
\usepackage{cleveref}
\crefname{equation}{equation}{equations}
\Crefname{equation}{Equation}{Equations}
\crefname{inequality}{inequality}{inequalities}
\Crefname{inequality}{Inequality}{Inequalities}
\creflabelformat{inequality}{(#2#1#3)}
\crefrangelabelformat{inequality}{(#3#1#4)--(#5#2#6)}

\makeatletter
\def\ps@titlepage{\let\@mkboth\@gobbletwo
  \def\@oddhead{}\def\@evenhead{}%
  \def\@oddfoot{\hfill\thepage\hfill}%
  \let\@evenfoot\@oddfoot}
\def\ps@headings{\let\@mkboth\markboth
  \def\@oddhead{\hfill{\itshape\@righttitle}\hfill}%
  \def\@evenhead{\hfill{\itshape\@lefttitle}\hfill}%
  \def\@oddfoot{\hfill\thepage\hfill}\let\@evenfoot\@oddfoot}
\makeatother
\newcommand\cupnotice{This article has been accepted for publication in a revised form in
  \emph{Journal of Fluid Mechanics},
  \url{https://www.cambridge.org/core/journals/journal-of-fluid-mechanics}.
  This version is free to view and download for private research and study only.
  Not for re-distribution or re-use. \copyright\ The Author(s), 2026.}

\title{A Steady Theory for Highly Unsteady Thin Airfoil Motions}

\righttitle{A Steady Theory for Highly Unsteady Thin Airfoil Motions}
\lefttitle{S. A. Bollt}

\author{Scott A. Bollt\aff{1}}

\affiliation{\aff{1}Department of Aerospace, California Institute of Technology, Pasadena, CA 91125, USA}

\corresau{Scott A. Bollt, \email{sbollt@caltech.edu}}

\begin{document}

\maketitle
{\renewcommand\thefootnote{}\footnotetext{\cupnotice}}

\begin{abstract}
A generalized steady thin airfoil theory is built not from steady flow, but from flows in which there exists a frame of reference where the absolute vorticity field is steady. We specialize this general idea to construct a foliation of a domain of phase space for thin airfoil motions in which staying on a leaf of the foliation maintains zero absolute vorticity in the fluid indefinitely. The phase space is described by the airfoil's total rotation rate, speed, and angle of attack. It is found using this definition that one of the preeminent unsteady stall conditions is equivalent to the traditional steady stall condition for this class of flows, various other classical results can be re-interpreted as steady in the absolute vorticity field, and insight is also provided into the surprising experimental results of \citet{mcgowan2011investigations}. We show parts of existing quasi-steady flow and Theodorsen unsteady flow theory can be built from a planar approximation of our foliation, making the quasi-steady extension of our theory trivial, and suggesting how an unsteady extension can be created as well. Furthermore, pure unsteady effects in the new theory amount to a transversal of the leaves, occupying just one dimension, inviting interesting prospects for future experimental and theoretical work.
\end{abstract}

\section{Introduction}
\label{sec:ch5-background}
Despite receiving substantial attention, unsteady airfoil aerodynamics is still a fruitful topic of study with many unsolved problems. These problems play central roles in emerging applications such as unmanned aerial vehicle control, enhanced-maneuverability aircraft, gust response, aeroelasticity, and cycloidal propellers and wind turbines. They are also central to questions of biological locomotion in fluid environments.
 
A key phenomenon discovered experimentally, known as the Kramer effect \citep{kramer1932zunahme}, is that when a flat plate or thin wing is pitched while the wing translates in a straight line, the pitching behavior can keep the flow attached beyond the angle of attack at which stall normally occurs. When the pitching rate is high, the achievable angle can become much higher than the steady stall angle \citep{strickland1986dynamic,Carr1988}. This forms the basis of supermaneuverability \citep{Herbst1983}. Rapid pitching also partially explains the high lift coefficients observed in insect flight \citep{Ellington1996,Dickinson1999}. 
 
The body of unsteady aerodynamic study is focused on understanding flow response to gusts and to canonical airfoil motions: pitch, heave, and surge \citep{Eldredge2019rev}. For fast-moving aircraft these motion primitives are natural because the dominant motion is straight flight at constant speed and angle of attack, with any deviation being small (e.g.\ aircraft pitch when taking off, and they heave in gusts). The pitch-heave-surge decomposition also happens to be a standard Cartesian representation. However, it is not obvious that this is the most natural way to describe unsteady airfoil motion. When a bird perches, or an insect hovers, or a drone encounters a gust whose strength is comparable to its forward speed, the fluid dynamics becomes more complex. In those situations it is useful to use a coordinate system to describe airfoil motion which is intrinsic to the vortex dynamics, not Cartesian coordinates.

Modern steady lift theory arguably began with the work of Lanchester, and Kutta's study of a circular arc \citep{kutta1902auftriebskrafte}. The circular arc's shape avoided many of the philosophical hurdles which had hindered prior progress since it has a unique, singularity-free, and lift-generating flow solution. Recognition that airfoil leading edges are typically rounded later resolved the issue regarding flat plate leading edge singularities \citep{chaplygin1910pressure,kutta1910grundlagen}: a singularity in the flat plate model is acceptable so long as the flow around the airfoil's rounded leading edge is not too adverse for the boundary layer to navigate, which is true below some critical stall angle of attack.

Unsteady thin airfoil theory is more complex than its steady counterpart. Classical work on pitching, heaving, and gust response \citep{Wagner1925,Theodorsen1934,VONKARMAN1938} allows treatment of flows due to mildly unsteady airfoil motion, but not highly unsteady motion. Inviscid simulations allow expedient modeling of flow over sharp-edged bodies well beyond the point of stall, like a flat plate at high angle of attack, by assuming point vortices are shed at both edges so as to satisfy the Kutta condition at all times \citep{Sarpkaya1975}. More recently, to tackle thin airfoils crossing through the stall regime, the concept of the critical leading edge suction parameter (LESP) was introduced by \citet{ramesh2011augmentation}. This model assumes that the Kutta condition applies to the trailing edge, but at the leading edge the flow remains attached so long as the leading edge suction parameter (i.e.\ the leading edge singularity's strength) does not exceed a critical value. Once the critical value is reached, vorticity emanates from the leading edge. It was realized by those authors that the critical value depends on the specifics of the airfoil, but is not very sensitive to the airfoil's motion (although treating the value as a true variable does offer some further modeling improvements \citep{Darakananda2018}). One can recognize this LESP as a generalization, applicable to aggressive airfoil motion, of the same philosophical insight which was introduced by \citet{kutta1910grundlagen}: ``I consider the application of the [lift] formulae to large angles of attack... to be highly suspect physically. ...the actual processes will still be complicated by vortices generated at the leading edge, especially if there is insufficient rounding there.''\footnote{English translation provided by \citep{Axcell2001}} The LESP formulation is conceptually identical, with \citet{kutta1910grundlagen} even justifying leading edge rounding via asymptotic expansion akin to \citet{ramesh2020leading}. The primary underlying conceptual difference appears to come from whether the leading edge suction depends on time. Nevertheless, the idea has recently allowed impressive capabilities to emerge, like the inverse design of airfoil kinematics to achieve goals such as unsteady lift-coefficient tracking \citep{suresh2024lift}.

Studies of the self-similar vortical flow produced by sharp edges offer insights into how suppression of vorticity shedding can be achieved. Point-vortex flow due to motion parallel and perpendicular to an edge was considered by \citet{Cortelezzi1993}, and it was found there are certain motions which make vortex shedding inadmissible. \Citet{Pullin2020} discovered that it is possible to suppress generation of a vortex sheet at the trailing edge of a flat plate for small time. Recently it was discovered that the same phenomenon can occur at the leading edge too \citep{Hinton2024}, and the results have been validated with direct numerical simulations \citep{sader2024starting}. While these theories are built around small-time asymptotic analysis, the underlying reason vorticity generation is suppressed carries no time restriction. Furthermore, it should be recognized that the resulting flow has a time-invariant absolute vorticity field (the vorticity field using rotation rate with respect to an inertial frame) by construction.

If the basis of modern unsteady models and theories is actually a subtle (though crucial) generalization of the earliest steady thin airfoil theories, it calls into question whether we are tracking the right quantity when we build steady and unsteady airfoil theories. Concern only grows in view of results like \citet{mcgowan2011investigations} wherein rapid and large combined pitch-heave motions which produce nearly constant force can be constructed from theories only meant to handle small and slow pitch and heave. In particular, these observations suggest that the important quantity that should be steady is the absolute vorticity field in any reference frame, not the flow. We apply this concept to the case of thin airfoil theory and specializing to the case of a trivial absolute vorticity field. This results in a requirement that the airfoil move on a single 2D leaf within a 3D phase space. This perspective may allow existing quasi-steady and unsteady theories to be reapplied to flows where airfoil motion is highly unsteady but the absolute vorticity field yields to perturbation analysis.

\section{Unsteady Thin Airfoil Modeling}
\label{potential flow section}
In this section we summarize a few well-established results from potential flow theory for unsteady thin symmetric airfoil motion. We assume, to simplify the analysis, that the fluid at infinity is not moving in the ground frame, but the results herein can all be easily generalized to cases in which that is not the case by constructing an intermediate wind frame.

\begin{figure}[ht!]
    \centering
        \centering
        \includegraphics[height=2.45in]{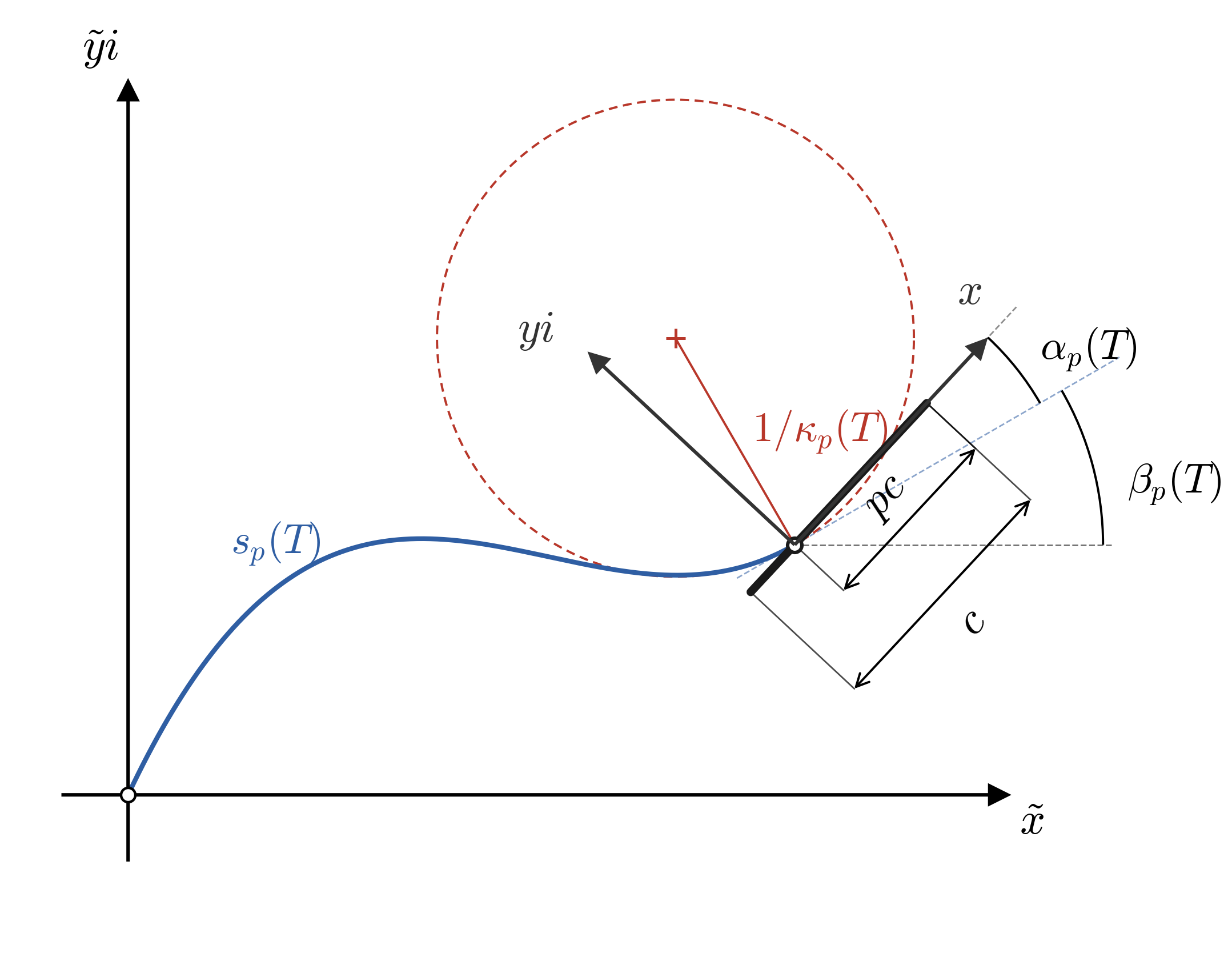}
        \caption{Path-coordinate representation of the thin airfoil's kinematics. The airfoil (black segment) has chord length $c$. The center of rotation ($\circ$) is the origin of the body-fixed $z$ plane and is located a constant distance $pc$ from the airfoil's leading edge. The trajectory of the center of rotation (blue curve) is parameterized by its arclength, $s_p(T)$, and its slope, $\beta_p(T)$. The airfoil's orientation is parameterized by angle of attack, $\alpha_p(T)$. The osculating circle of the trajectory at the center of rotation (red dashed curve, center marked $+$) has radius $1/\kappa_p(T)$. Code to generate figure written in Python, assisted by Claude Code (Opus 4.7 and 4.8 LLM).}
        \label{dynamics figure}
\end{figure}

Consider an airfoil with zero thickness and camber, and chord length $c$, as depicted in \cref{dynamics figure}. Such an airfoil is typically modeled by a flat plate. The ground frame is the $\tilde{z}=\tilde{x}+\tilde{y}i$ plane. We parameterize the airfoil's motion in path coordinates, where $s_p(T)$ is the distance along the airfoil's path to its current location, $\beta_p(T)$ is its instantaneous angular direction of motion with respect to the lab frame axes, and $\alpha_p(T)$ is the instantaneous angle of attack between the airfoil and its direction of motion. We are using the dimensionless distance-traveled $T\equiv s_p/c$ to parameterize the trajectory rather than clock-time. Zero background flow is assumed. Consider any rotation of the airfoil to occur about a local point of rotation, $p$, where $p=1$ is the trailing edge of the airfoil, and $p=0$ is the leading edge. The subscript $p$ has been used in the motion parameterization because the parameterization depends on the point of rotation used to describe the motion, even for identical motions. This subscript will be used for other motion parameters for the same reason. Another coordinate system (complex form: $z=x+yi$) is centered at the airfoil's point of rotation and is aligned with the airfoil. The body-fixed plane conformally maps to the unit circle in the $\zeta$-plane by
\begin{equation}
    \frac{z}{c}=\frac{1}{4}\left(\zeta+\frac{1}{\zeta}\right)+p-\frac{1}{2}.
    \label{eta}
\end{equation}

In \citet[\S\,9.65, equations (4) and (5)]{Milne-Thomson} the ground frame potential flow solution for an elliptic cylinder of major and minor diameters $2a$ and $2b$ is provided. They use a coordinate system where the major axis lies along the $x$-axis. Flow approaches the ellipse at speed $U$, and angle $\alpha_{M-T}$ with respect to the $x$-axis. The ellipse's instantaneous rotation rate about its center is given as $\omega_{M-T}$. A flat plate results if we set $b=0$. They write the complex potential as,
\begin{equation}
    f(z)=\frac{iUa \sin{\alpha_{M-T}}}{\zeta}+\frac{ia^2\omega_{M-T}}{4\zeta^2}.
\end{equation}
We can express this in the body frame by adding in the flows which appear due to the frame-change, which are a uniform flow and rigid body rotation flow,
\begin{equation}
    f(z)=Ue^{-i\alpha_{M-T}}z+\frac{iUa \sin{\alpha_{M-T}}}{\zeta}+\frac{ia^2\omega_{M-T}}{4\zeta^2}-\frac{i\omega_{M-T}}{2} \overline{z}z.
\end{equation}
We can use this solution to determine the potential flow for general rigid-body motion described in path coordinates. From \cref{dynamics figure}, $\alpha_{M-T}\to -\alpha_p(T)$, since in our figure $\alpha_p(T)$ is the angle from the airfoil's direction of motion to the $x$-axis, as measured from the rotation point $p$. The total angular rate of the airfoil's motion is the sum of the rate of change of angle of attack and its heading direction, so $\omega_{M-T}\to \dot{\alpha}_p(T)+\dot{\beta}_p(T)$. This sum appears frequently, so we represent it by $\Omega(T)$. The airfoil's speed is $\dot{s}_p(T)$ so $U\to \dot{s}_p(T)$, and the airfoil has chord $c$ rather than $2a$. They are also using an old convention for the complex potential which is the same up to multiplication by $-1$. Finally, the speed of the center of the airfoil normal to its surface was $U\sin(\alpha_{M-T})$ in \citet{Milne-Thomson}, but for us the center of rotation can be offset from the airfoil's center, so the speed is $c(\dot{\alpha}_p(T)+\dot{\beta}_p(T))(1-2p)/2+\dot{s}_p(T)\sin(\alpha_p(T))$. These modifications, along with allowing the airfoil to have non-zero circulation, $\gamma(T)$, implying a starting vortex was shed into the far-field, yield an expression for the complex velocity (in dimensionless form),
\begin{equation}
    A(T)=i\left[-\kappa_p(T)(p-1/2)+\sin(\alpha_p(T))\right],
    \label{A path}
\end{equation}
\begin{equation}
    B(T)=-i\kappa_p(T)/8,
    \label{B path}
\end{equation}
\begin{equation}
    \frac{\overline{u}(\mathbb{z},T)}{\dot{s}_p(T)}=-e^{i\alpha_p(T)}+i\kappa_p(T)\overline{\mathbb{z}}-2\frac{A(T)+2\frac{B(T)}{\zeta}}{\zeta^2-1}-\frac{2\Gamma(T)i}{\pi \mathbb{U}_p(T)}\frac{\zeta}{\zeta^2-1}.
    \label{discrete vortex model infinite trailing edge}
\end{equation}
In the above expression we have introduced definitions for a few dimensionless quantities. They are dimensionless airfoil speed of motion, $\mathbb{U}_p(T)\equiv \frac{\dot{s}_p(T)}{\dot{s}_0}$ ($\dot{s}_0$ is some normalizing speed, such as the speed of the airfoil when a given circulation was acquired), dimensionless airfoil curvature of motion, $\kappa_p(T)\equiv\frac{c\Omega(T)}{\dot{s}_p(T)}$, dimensionless circulation, $\Gamma(T)=\frac{\gamma(T)}{c\dot{s}_0}$, and the chord-normalized complex plane, $\mathbb{z}\equiv \frac{z}{c}$.

\subsection{Vorticity Generation Models}
\label{subsec:ch5-the-kutta-condition-and-lesp}
At the trailing edge, the Kutta condition demands no singularity can exist in the flow, and vorticity will be shed by that edge whenever required to enforce this. Everywhere else on the airfoil's surface, we choose to focus on models of vorticity shedding which are related directly to the airfoil's instantaneous state of motion. The details of these will be covered in this section.

Applying the Kutta condition requires knowing the flow's surface velocity. It can be found by representing the body's surface in the transformed coordinates using $\zeta=e^{i\theta}$ (a useful result is that on the surface, $\mathbb{z}=\frac{1}{2}(\cos{\theta}-1+2p)$), then applying these definitions to \cref{discrete vortex model infinite trailing edge}. The result is,
\begin{equation}
    \frac{u_s(\theta,T)}{\dot{s}_p(T)}=\frac{-\frac{\kappa_p(T)}{2}((1-2p)\cos(\theta)-1/2\cos(2\theta))-\sin(\alpha_p(T)+\theta)-\frac{\Gamma(T)}{\pi \mathbb{U}_p(T)}}{\sin(\theta)}.
    \label{surface flow equation}
\end{equation}
From \cref{surface flow equation}, the Kutta condition requires the numerator to be zero at the trailing edge ($\theta=\pi$) such that the denominator does not produce a singularity. The Kutta condition thus yields,
\begin{equation}
    (p-3/4)\kappa_p(T)=\sin(\alpha_p(T))-\frac{\Gamma(T)}{\pi}\frac{1}{\mathbb{U}_p(T)}.
    \label{dimensionless TEV constraint equation}
\end{equation}

Shedding of vorticity elsewhere on an airfoil has been modeled in a variety of ways historically. For us the important thing is that we use a model which is a memoryless function of the airfoil's instantaneous state of motion, 
\begin{equation}
g(\mathbb{U}_p,\alpha_p,\kappa_p)\leq 0.
\label{general shedding inequality}
\end{equation} 
Notice that when we use \cref{dimensionless TEV constraint equation} to solve for $\Gamma(T)$, we can use the result to obtain a form of \cref{discrete vortex model infinite trailing edge} which looks like $\overline{u}(\mathbb{z},T)/\dot{s}_p(T)=h(\mathbb{U}_p,\alpha_p,\kappa_p)$, so any vorticity shedding criterion which depends on only $\overline{u}(\mathbb{z},T)/\dot{s}_p(T)$ (a.k.a. the boundary layer is itself a state-function of the instantaneous outer-flow), and does not shed at all for some subset of outer-flows, is compatible with inequality~(\ref{general shedding inequality}) automatically. Examples include steady boundary layer separation criteria (critical angle of attack, Stratford separation \citep{stratford1959prediction}, Head's method \citep{Head1958}, etc.), as well as unsteady leading edge shedding criteria like critical LESP \citep{ramesh2011augmentation}.

We will default to the critical LESP model as a specific realization of inequality~(\ref{general shedding inequality}) to carry forward certain calculations. It is a premier model for unsteady separation on thin airfoils. The idea is that at the airfoil leading edge, $\theta=0$, so the flow velocity is typically unbounded under a flat-plate model and thus even in the presence of slight rounding, separation is likely to occur there if it happens at all. The LESP is defined as the strength of the singularity in \cref{surface flow equation}: $\mathbb{S}_p(T)\equiv\lim_{\theta\to 0}\frac{\sin{\frac{\theta}{2}}}{2} \frac{u_s(\theta,T)-u_s(-\theta,T)}{U}$, for some normalizing speed $U$. The critical LESP is a chosen value beyond which vortex shedding is predicted to occur \citep{ramesh2011augmentation}. Because this model only cares about the strength of the leading edge flow singularity it is particularly simple: it is local, not just memoryless. According to \citet{Ramesh2014}, the value of the threshold is nearly independent of airfoil motion, though not entirely \citep{Narsipur2020}. Mathematically, the criterion is,
\begin{equation}
    |\mathbb{S}_p(T)|<\mathbb{S}_{crit}.
    \label{LESP criticality}
\end{equation}
Applying the LESP definition to \cref{surface flow equation} we obtain,
\begin{equation}
    (p-1/4)\kappa_p(T)=\sin(\alpha_p(T))+\frac{\Gamma}{\pi}\frac{1}{\mathbb{U}_p(T)}+2\frac{U}{\dot{s}_p(T)}\mathbb{S}_p(T).
    \label{dimensionless LEV suppression condition general}
\end{equation}
Summing \cref{dimensionless LEV suppression condition general} and \cref{dimensionless TEV constraint equation} yields a version with circulation eliminated,
\begin{equation}
(p-1/2)\kappa_p(T)=\sin(\alpha_p(T))+\frac{U}{\dot{s}_p(T)}\mathbb{S}_p(T).
\label{general null sum}
\end{equation}

\section{Steady Absolute Vorticity Fields and Unsteady Airfoil Motion}
Consider the flow around a 2D body moving in a general manner in which the absolute vorticity field is time-invariant in some frame. Changing reference frames does not change the physics which are taking place, and \emph{the absolute vorticity equation is invariant to every possible change of Euclidean reference frame} \citep{Eldredge2019rev}. Therefore, if any reference frame exists which makes the absolute vorticity field time invariant, the same field can be recreated in a scenario where the steady frame of reference is itself a fixed frame; the absolute vorticity field can be steady even when airfoil motion is highly unsteady. For the particular case studied here, the special frame is the body-fixed frame, and the equivalent flow is one in which the airfoil is fixed in place and the fluid is made to, at each point in time, spin in rigid body rotation at a rate $-\Omega(T)$ about a point $\dot{s}_p(T)/\Omega(T)$ from the point $p$ on the airfoil where this point is at an angle $\pi/2-\alpha_p(T)$ from the (body-fixed) $x$-axis. The absolute vorticity field's dynamics are identical for the flow of interest and this dual-flow, since they are identical up to a change in frame, but clearly the absolute vorticity field of the constructed dual-flow is steady if the field does not change in time. Hence, we should consider the absolute vorticity field of the flow generated by the moving and accelerating airfoil to also be steady.

As seen in \citet{VONKARMAN1938}, it is not the assumption of a steady airfoil motion or steady flow, or even a flat wake, which is fundamental to their unsteady theory, but a vortical wake in which all vorticity lies on a one-dimensional curve whose ground-frame motion is zero. Thus, the most general steady airfoil theory is not one which restricts itself to steady airfoil motion, but instead with steady absolute vorticity in a carefully chosen frame.

\section{A Foliation for the Thin Airfoil Motions Which Produce Steady Absolute Vorticity Fields}
\label{vortex suppression section}
For a thin symmetric airfoil, the absolute vorticity field cannot be steady in the presence of vortex shedding, so \cref{dimensionless TEV constraint equation} and inequality~(\ref{general shedding inequality}) must be treated not as vortex shedding models but constraints on rigid body motion; rotate or accelerate the airfoil too slowly and vorticity is shed, but too fast and vorticity of the opposite sign is shed. While our analysis is for irrotational flow and thin symmetric airfoils, this idea is clearly more general. 

Applied to the flow model, we see \cref{dimensionless TEV constraint equation} defines a family of embeddings of non-compact two-dimensional manifolds, parameterized by $\Gamma$, in three-dimensional phase space (coordinates $(\alpha_p,\kappa_p,\mathbb{U}_p)$). Meanwhile, inequality~(\ref{general shedding inequality}) defines a fixed volume in the same phase space in which vortex shedding everywhere else is also avoided. Each manifold, $\mathcal{M}_\Gamma$, is defined to be the intersection of the shedding-free volume with the 2D embedding at a given $\Gamma$. A leaf, $\mathcal{L}_{\Gamma,j} \subseteq \mathcal{M}_\Gamma$, is a connected component of one of these manifolds. The set of all leaves is a foliation of the domain carved out by inequality~(\ref{general shedding inequality}), $\mathcal{S}=\{\mathcal{L}_{\Gamma,j}\}_{\Gamma\in\mathbb{R},j\in \mathbb{N}}$.

Any airfoil trajectory, $\eta(T):\mathbb{R}\to\mathbb{R}^3$, for which all of its states lie on a single leaf, $\eta\subset\mathcal{L}_{\Gamma,j}$, does not shed vorticity at all. We can compare $\mathcal{S}$, and the domain it foliates, to the region of phase space existing steady airfoil theory applies to. Each leaf, $\mathcal{L}_{\Gamma,j}$, is 2D and lives in a 3D space, whereas only pitch-free, heave-free, and surge-free motion is considered by existing steady theory: a single point in 3D phase space. Only trajectories which traverse leaves shed vorticity. The leaf transversal is parameterized by $\Gamma$ and is just one-dimensional.

\subsection{Inequality~(\ref{general shedding inequality}) Under a Critical LESP Model}
Among the various choices one can make for inequality~(\ref{general shedding inequality}), the critical LESP model is particularly interesting because whenever the airfoil's motion stays on a single leaf, the LESP model collapses onto the steady stall angle of attack. A natural consequence is then that each manifold is composed of a single leaf for realistic stall angles because the embedding is injective over the allowed range of $\alpha_p$ and the allowed range is a connected interval. As a consequence, $\mathcal{M}_\Gamma=\mathcal{L}_{\Gamma,j}$ in this model. 

To show the collapse of the critical LESP model into a steady stall condition we first choose to define the normalizing speed in the LESP as $U\equiv \dot{s}_p(T)$ to eliminate the $U/\dot{s}_p(T)$ factor in \cref{general null sum}. The second choice is to make $p=1/2$. Any airfoil motion can be described by a combination of translation and rotation about any instantaneous center of rotation so this too is something we are free to choose without introducing restrictions on the motion which can be described. The result of these two definitional choices is that \cref{general null sum} becomes,
\begin{equation}
\mathbb{S}_{1/2}(T)=-\sin(\alpha_{1/2}(T)),
\label{angle suction relation}
\end{equation}
without loss of generality. Applying this formula to inequality~(\ref{LESP criticality}) results in,
\begin{equation}
|\alpha_{1/2}(T)|< \alpha_{ss}.
\label{critical angle inequality}
\end{equation} 
It has already been claimed that the LESP can be interpreted as an ``\,`effective unsteady angle of attack' consisting of contributions from the geometric
angle of attack, aerofoil camber, plunge velocity and induced velocities by the wake'' \citep{ramesh2018leading}. By utilizing $\dot{s}_{p}(T)$ to define $\mathbb{S}_{p}(T)$ and choosing $p=1/2$ the connection goes much further: the critical LESP loses any distinction from the static stall angle of attack, $\alpha_{ss}$, for thin airfoils moving on $\mathcal{M}_\Gamma$, because in that case inequality~(\ref{critical angle inequality}) applies. Inequality~(\ref{critical angle inequality}) has no way to distinguish between an airfoil moving without pitching, surging, or plunging, and more general motions.

\subsection{Visualizing $\mathcal{M}_\Gamma$}
\begin{figure}[ht!]
    \centering
    \begin{subfigure}[b]{0.44\textwidth}
        \centering
        \includegraphics[width=\textwidth]{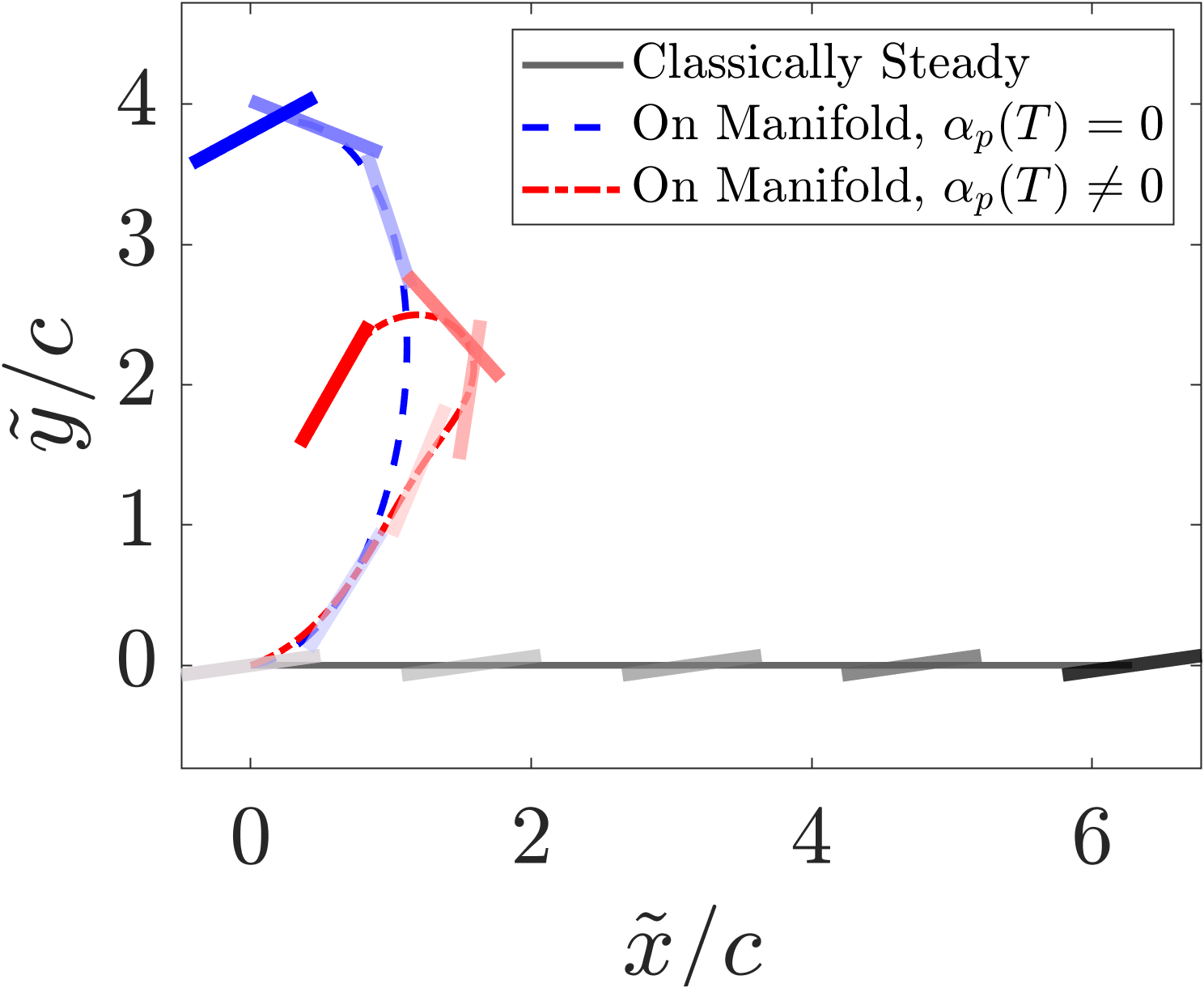}
        \caption{The example phase-space trajectories used in \cref{fig:generalized steady trajectories phase}, shown as physical airfoil motion in the ground frame. For each we show the airfoil at various points in time (older snapshots are more transparent), together with its trajectory (thin curves).}
        \label{fig:generalized steady trajectories physical}
    \end{subfigure}
    \hfill
    \begin{subfigure}[b]{0.55\textwidth}
        \centering
        \includegraphics[width=\textwidth]{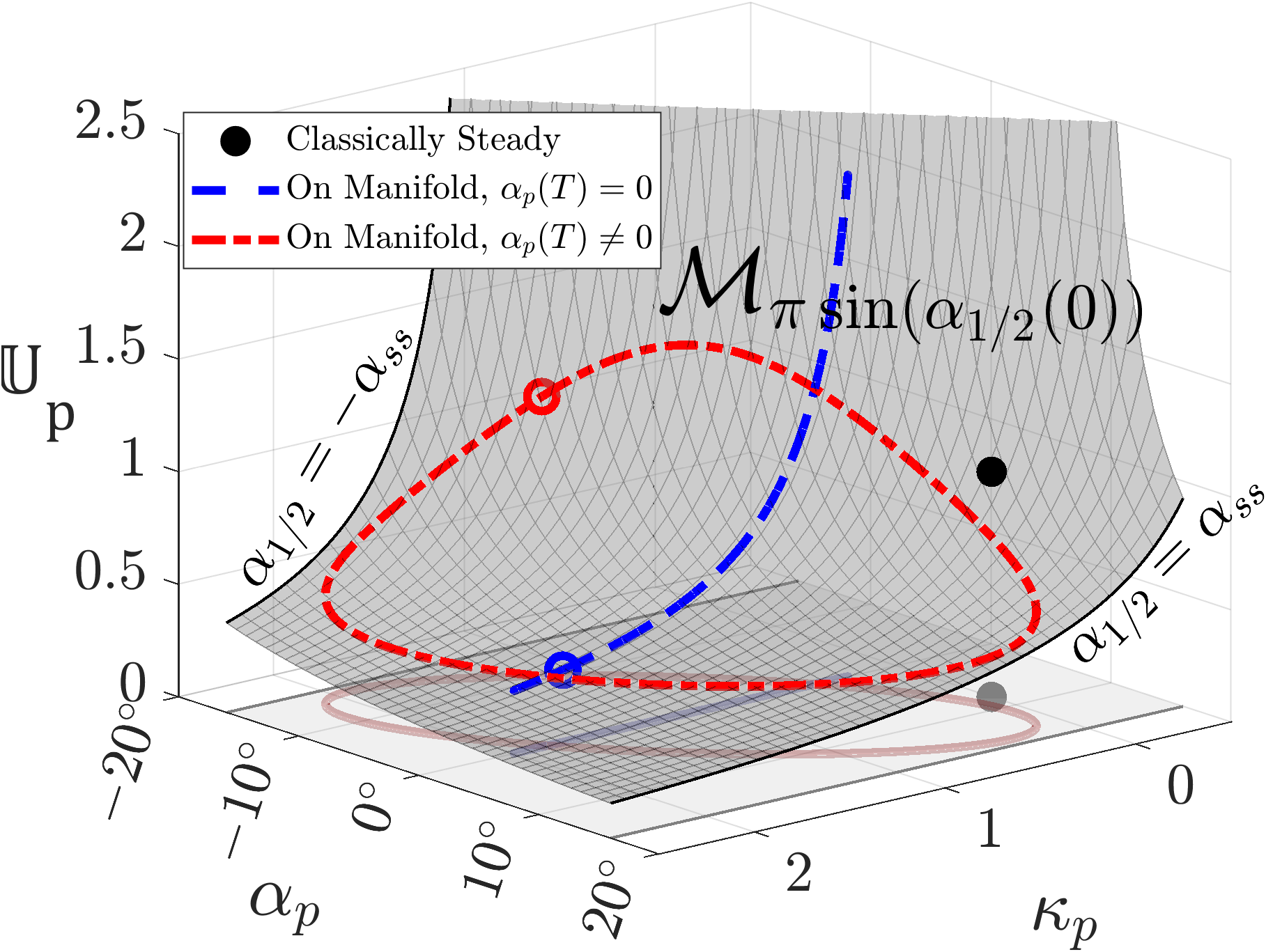}
        \caption{$\mathcal{M}_{\pi\sin\alpha_{1/2}(0)}$ as viewed in the phase space formed by $(\kappa_{1/2},\alpha_{1/2},\mathbb{U}_{1/2})$. A representation of the trajectories from \cref{fig:generalized steady trajectories physical} in this space is also shown. (gray surface) The manifold, $\mathcal{M}_{\pi\sin\alpha_{1/2}(0)}$. Each trajectory starts at the black dot. The classical steady trajectory must stay there, while the others jump to the respective locations marked with an open circle and proceed to traverse the shown elliptical path in $(\alpha_{1/2},\kappa_{1/2})$ space.}
        \label{fig:generalized steady trajectories phase}
    \end{subfigure}
    \caption{Trajectories that are on $\mathcal{M}_{\pi\sin\alpha_{1/2}(0)}$ in physical space and in phase space.}
    \label{fig:generalized steady trajectories}
\end{figure}

The surprising difference between classically steady trajectories and those which can have a steady absolute vorticity field is shown in \cref{fig:generalized steady trajectories physical}. This figure illustrates three trajectories which initially have the same straight-flight motion and common circulation, $\Gamma = \pi\sin\alpha_{1/2}(0)$, associated with classical steady thin airfoil theory. One example trajectory continues on this straight motion, the second stays on $\mathcal{M}_{\pi\sin\alpha_{1/2}(0)}$ with zero LESP, and the third is on $\mathcal{M}_{\pi\sin\alpha_{1/2}(0)}$ with non-zero LESP. 

The same trajectories as shown in \cref{fig:generalized steady trajectories physical} are provided in \cref{fig:generalized steady trajectories phase} as curves through phase space, with the steady manifold $\mathcal{M}_{\pi\sin\alpha_{1/2}(0)}$ also displayed. The open circles represent the point the example trajectories jump to at the start of the deviation from straight-line motion for the three examples which do so. The jump represents a discontinuity in the airfoil's velocity, not its position or orientation. Though such phase-space jumps don't break the inviscid model, the trajectory was chosen for its simplicity rather than its physical usefulness. Notice the classic steady trajectory (black) occupies a single state on $\mathcal{M}_{\pi\sin\alpha_{1/2}(0)}$ (the state the three trajectories all start at). One can easily show $\mathbb{U}_{1/2}$ is constant only along curves of constant $\kappa_{1/2}+4\sin(\alpha_{1/2})$.

Even if we set $\alpha_{ss}=0$ so that inequality~(\ref{general shedding inequality}) enforces the Kutta condition at the leading edge, $\mathcal{M}_\Gamma$ does not become a single point. Instead it becomes a 1D curve in the $(\kappa_p,\alpha_p,\mathbb{U}_p)$ subspace: $\alpha_p(T)=0$, and $\dot{\beta}_p(T)$ is solely determined by the constant circulation, but the airfoil's speed is completely free. The blue curve in \cref{fig:generalized steady trajectories} are an example of such a scenario. 

\section{Unsteady Forces From a Steady Absolute Vorticity Field}
\label{subsubsec:ch5-force}
Though the absolute vorticity field is steady, the forces experienced are necessarily not because the body can accelerate. As we will soon see, there is a clean split between force due to the steady absolute vorticity field and the unsteady boundary condition if $p=1/2$. We then show that our theory allows a re-casting of a few quasi-steady lift results as steady, and re-interpret some surprising aspects of a modern experiment. The formula we provide removes the small motion and time-derivative restrictions that typically come with quasi-steady and unsteady airfoil theories, while maintaining the exact construction from potential flow that is enjoyed by existing steady theory.

Force on a flat plate moving in an arbitrary manner is treated using vortex impulse by \citet{Eldredge2019}. Our notation and assumptions reduce their relation to,
\begin{multline}
    c_F(T) = \frac{i\pi}{2}\left[\frac{\mathbb{U}'_p(T)}{\mathbb{U}_p(T)}\sin(\alpha_p(T))+\cos(\alpha_p(T))\kappa_{\alpha_p}(T)-i\sin(\alpha_p(T))\kappa_p(T)\right]e^{-i\alpha_p(T)}\\
    -\frac{i\pi(p-1/2)}{2}\left(\frac{\mathbb{U}'_p(T)}{\mathbb{U}_p(T)}\kappa_p(T)+\kappa'_p(T)-i\kappa_p^2(T)\right)e^{-i\alpha_p(T)}+2(p-1/2)\frac{\Gamma\kappa_p(T)}{\mathbb{U}_p(T)}e^{-i\alpha_p(T)}+\frac{2\Gamma i}{\mathbb{U}_p(T)}
    \label{coefficient of force suppression general}
\end{multline}
where $c_F\equiv \frac{D +Li}{1/2\rho c \dot{s}_p^2}$ is a complex representation of the lift and drag coefficients, $T\equiv s_p(T)/c$ is a distance-traveled parameterization of time, $\mathbb{U}'_p\equiv d\mathbb{U}_p/dT$, and $\kappa_{\alpha_p}\equiv \frac{c\dot{\alpha}_p}{\dot{s}_p}$. The derivation is provided in \cref{Force Coefficient Derivation}. The first two long terms are the added mass, which are uniquely related to the airfoil's motion. The third term uniquely depends upon both the airfoil acceleration and the circulation. The final term is the Kutta-Joukowski force, which does not depend on the body's acceleration. When $p=1/2$ the second and third terms drop out.

If we focus on just the Kutta-Joukowski (4th) term and use \cref{dimensionless TEV constraint equation} to solve for the circulation we get the following expression for the force in terms of airfoil kinematics:
\begin{equation}
    c_{F_J}(T)\equiv\frac{2\Gamma i}{\mathbb{U}_p(T)}=\left(2\pi(3/4-p)\kappa_p(T)+2\pi\sin(\alpha_p(T))\right)i.
    \label{Kutta-Joukowski force}
\end{equation}
The second term in this equation is the usual result from steady thin airfoil theory. The first term is known as the Kramer effect or ``rotational lift'' in the gust response \citep{kramer1932zunahme} and insect flight communities \citep{ellington1984aerodynamics,Dickinson1999}. This term is precisely equivalent to the one first derived by \citet{Munk1925}, but conceptually Munk's theory predicts lift due to vorticity newly shed in a slow pitching rate approximation in level flight, while our theory holds for any pitching rate but only if the motion is on $\mathcal{M}_{\Gamma}$. The explanatory power of rotational lift for insect flight, which is well outside the quasi-steady regime, may be explained in part by the fact that it can be derived from our steady perspective without restrictions on the size of time derivatives, not just a quasi-steady one. Furthermore, since the Kramer effect has already been shown to work as a quasi-steady theory, which amounts to slow excursions from trajectories which are a single point in $\mathcal{M}_{\Gamma}$, \cref{Kutta-Joukowski force} may hold for more general trajectories which have a small component of the trajectory which crosses between the leaves of $\mathcal{S}$.

The $\kappa_p$ term in \cref{Kutta-Joukowski force} connects trajectory curvature to airfoil camber. Kutta's analysis of the flow over a circular arc reveals the lift it produces is related to its sagittal height \citep{Axcell2001}, $h$, by $c_{l_{arc}} = 4\pi \frac{h}{c}\approx \frac{\pi}{2 \mathbb{r}}$, where $\mathbb{r}\equiv \frac{r}{c}$ is the ratio of the radius of curvature of the circular arc to the chord of the arc. The first term in \cref{Kutta-Joukowski force} therefore establishes an analogy between camber and curved motion when we set $p=1/2$ because the lift of a circular arc in the limit of large radius of curvature is the same as the lift due to curving a camber-free airfoil's motion ($1/\kappa_p$ may then be interpreted as trajectory radius of curvature when $\dot{\alpha}_p=0$). The observation that trajectory curvature acts like camber is not new: \citet{migliore1980} introduced the concept of ``virtual camber'' for Darrieus wind turbine blades and \citet{xu2021} extended the idea to a composite quasi-steady effective angle of attack for combined pitch, plunge, and surge. However, existing concepts have been heuristic rather than being derived from the full potential flow solution due to general airfoil motion.

\subsection{Comparison to Experiment}
A further consequence of \cref{coefficient of force suppression general} is a re-interpretation of the simulations and experiments performed by \citet{mcgowan2011investigations} in which they executed pitch and heave, in some cases in combination to target a constant total lift. They found that when their prescribed motion was designed using quasi-steady thin airfoil theory (QSTAT), essentially no wake was shed in two cases, regardless of whether the reduced frequency is high (cases (1,D) and (4,D)). 

The authors correctly identified that no wake is shed for (1,D) and (4,D) because making the quasi-steady lift constant suppresses circulation production, not force production. QSTAT is effectively a steady-vorticity lift model which is missing added mass terms (the authors commented on the missing added mass effect). The shed circulation is correctly captured by QSTAT even at high frequencies because circulation is constant (case (4,D)). In \cref{Asymptotic Equivalence of Lift Formulas} we show that the quasi-steady formula they used is a small \emph{amplitude} Maclaurin series expansion of the Kutta-Joukowski lift component that comes from motion on $\mathcal{M}_\Gamma$, and thus their D-cases are first order approximations of the embedding of $\mathcal{M}_\Gamma$ itself.

\begin{figure}[ht!]
    \centering
    \includegraphics[width=\textwidth]{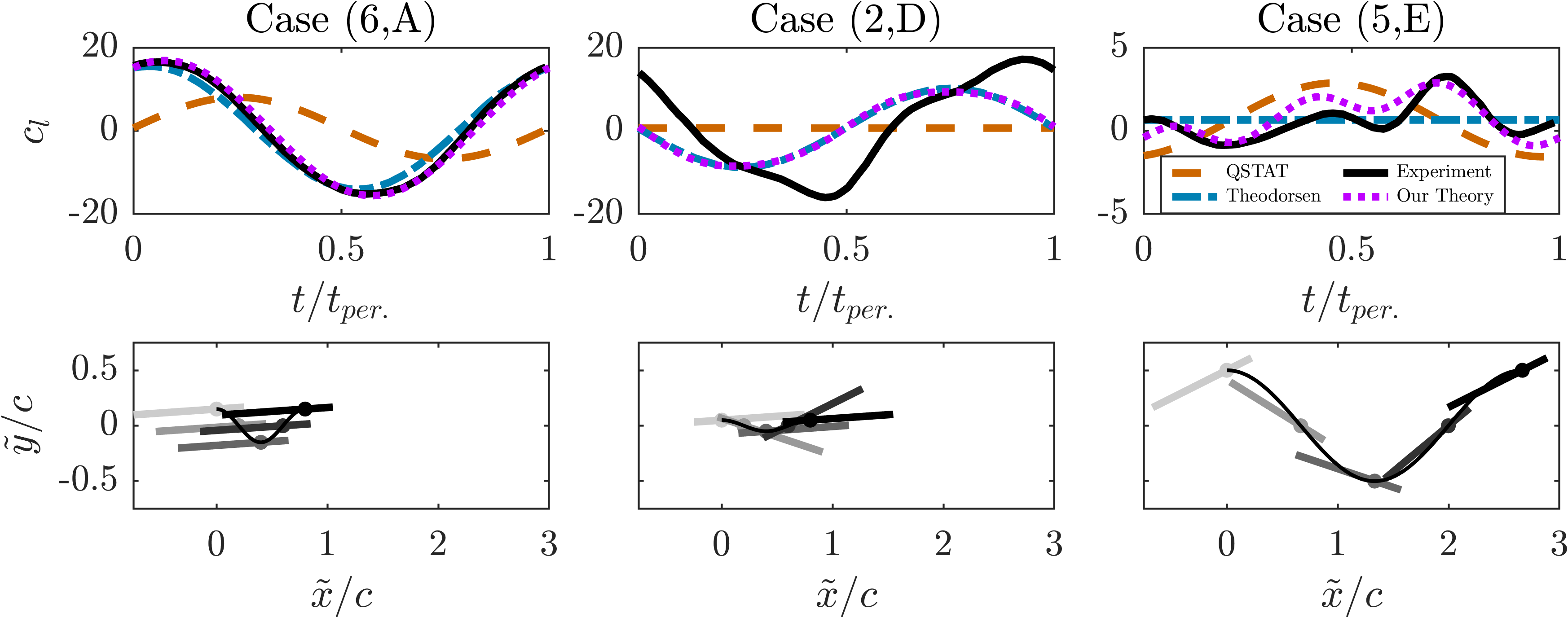}
    \caption{Lift coefficient over one period, $t_{per.}$, for three of the airfoil motions treated by \citet{mcgowan2011investigations} (top row), along with the wind frame motion of the airfoil showing its location at uniformly spaced snapshots in time (bottom row). Their experimental data, hand-digitized from their figures 20, 12 and 19, are the $-$ black curve; Theodorsen's theory is the $-\cdot$ blue curve, QSTAT is the $--$ orange curve, and our theory is the $\cdot\cdot$ magenta curve.}
    \label{fig:compare-to-experiment}
\end{figure}

In all four cases the maximum $\alpha_{1/2}$ in their experiment met or exceeded $\alpha_{ss}\approx 8^\circ$ for their airfoil \citep{galbraith2008implicit}: $\max(\alpha_{1/2})=15^\circ,40^\circ,8^\circ,15^\circ$, respectively for (1,D), (2,D), (3,D), and (4,D). Therefore, stall and leading edge vortex (LEV) development are a question of how long and by how much the critical angle is exceeded. In case (2,D) the stall angle was exceeded to such a degree that a powerful leading edge vortex was generated, whereas in cases (1,D) and (4,D) a small LEV is visible but its growth apparently gets cut off too quickly for it to become a dynamically important feature. In the middle column of \cref{fig:compare-to-experiment} we show the lift coefficient measured by the authors alongside our theoretical lift predictions and a wind frame depiction of airfoil motion for (2,D). We see that Theodorsen's model and ours agree closely with each other but do not capture the effect of the LEV. Meanwhile, case (3,D)'s periodic motion is much slower than the others, slow enough that the flow had time to separate in their experiments. 

The authors were able to use Theodorsen theory to cancel a large fraction of the lift variation out to both rates \emph{and} amplitudes that broke the model's stated assumptions, especially when ``the noncirculatory lift contribution becomes relatively larger.'' A partial explanation is that, as we show in \cref{Asymptotic Equivalence of Lift Formulas}, the Theodorsen lift model is a first order expansion of the entire lift component of \cref{coefficient of force suppression general}, in small amplitude, at least when the reduced frequency is zero, and since their motions were constructed to minimize lift variations they were constructing motions close to $\mathcal{M}_\Gamma$. Consider now the first column of \cref{fig:compare-to-experiment}, case (6,A), where a rapidly heaving airfoil's lift force is predicted by our theory comparably well to Theodorsen's theory, or the last column of the same figure, case (5,E), where our theory predicts the shape and timing of large measured lift fluctuations but Theodorsen's theory predicts constant lift. Because these two cases cross leaves, we use a quasi-steady extension of our theory: \cref{coefficient of force suppression general} is applied in spite of leaf traversal, using the circulation determined at each instant by \cref{dimensionless TEV constraint equation}. Apparently Theodorsen's small-amplitude approximations are more of a liability in comparisons to our theory than its frequency-dependent wake handling is an asset, since our wakeless theory is more faithful to the experimental measurement in case (5,E) but no better in case (6,A). A further consequence is that much of the improvement of Theodorsen's theory over QSTAT in these cases (and others in their experiments, not shown) comes mostly from accounting for added mass, not accounting for the wake.

In light of these observations, which we interpret as a statement that small unsteady excursions from a planar approximation of $\mathcal{M}_\Gamma$ can be handled with a surprising degree of robustness using lift formulas built from perturbing about this same approximation, one is inclined to ask if general periodic excursions from $\mathcal{M}_\Gamma$ can be handled using models built out of perturbations about \cref{coefficient of force suppression general}. 

\section{Conclusions}
\label{sec:ch5-conclusions}
A generalized definition of steady thin airfoil theory is provided, which uses the existence of a frame in which the absolute vorticity field is steady. A thin symmetric airfoil is modeled as a flat plate with circulation, in an inviscid fluid.

We find a dimensionless 3D phase space for airfoil motion in which a volume of no absolute vorticity production is foliated. The leaves of the foliation are a set of connected 2D sub-manifolds, each parameterized by the airfoil's circulation. The critical LESP model of leading edge vortex shedding is shown to be equivalent to a steady stall angle of attack for any motion on a single leaf.

Traditionally, unsteady aerodynamics is probed via studies of canonical Cartesian motions: pitching, heaving, and surging. This work suggests studying unsteady aerodynamics with respect to the foliation: probing motion on the leaves to improve the shedding model by isolating boundary layer effects, as well as controlled transversal curves which cross leaves to isolate purely unsteady vortex wake effects.

While aspects of this theory are already validated by equivalences to existing theory and experiment, further tests of the theory using experiments will require care. Thin airfoil theory only applies strictly when the boundary layer itself becomes thin at sufficiently high Reynolds numbers, and this theory comes with the further challenge of applying to motion with high acceleration and kinematics which cannot be reached without special hardware. With that in mind it may be fruitful for early tests to use various boundary layer theories applied to the pressure fields obtained using a 2D potential flow solver. 

Quasi-steady and unsteady theories built from ours may be able to predict forces when crossing slowly between leaves, a great expansion on current theories such as Theodorsen's. Such extensions only require the unsteadiness to handle the 1D leaf transversal in order to connect the leaves across the entire foliation, suggesting even large deviations could potentially be probed experimentally since the space is reduced. Since the flow is non-trivial but free from shed vorticity, it could be used to benchmark inviscid numerical schemes, separating out their attached and vortical flow modeling capabilities. Changes to the foliation structure due to airfoil camber, time invariant vorticity fields, and even time-varying vorticity fields, are all natural extensions worth further consideration.

\backsection[Acknowledgements]{The author would like to thank John Sader, Dale Pullin, Anthony Leonard, and Tim Colonius for helpful discussions.

Use of AI, when relevant, described in figure captions.}

\backsection[Funding]{None.}

\backsection[Declaration of interests]{{\bf Declaration of Interests}. The author reports no conflict of interest.}

\bibliographystyle{jfm}
\bibliography{references}
\appendix
\section{Force Coefficient Derivation}
\label{Force Coefficient Derivation}
Here we derive an expression for the force on a symmetric thin airfoil with fixed circulation executing arbitrary motion. The force on a body moving in a fluid can be calculated from the kinematics of the vorticity field. In 2D an elegant expression in terms of complex variables becomes possible \citep[\S\,6.2.2, equation 6.48]{Eldredge2019}, and further simplifications occur with a flat-plate model \citep[\S\,8.2.1]{Eldredge2019}. If we modify the flow model from \cref{potential flow section} to allow the trailing edge vortex to be at finite distance from the plate, $\zeta_\Gamma(T)$, the impulse referenced to the ground frame is,
\begin{equation}
    I(T) = \frac{ic}{4}e^{i(\beta_p(T)+\alpha_p(T))}\left[-\pi c\dot{s}_p(T)\sin(\alpha_p(T))+\pi c\dot{s}_p(T)(p-1/2)\kappa_p(T)+\gamma\left(\zeta_\Gamma(T)-\frac{1}{\overline{\zeta}_\Gamma(T)}\right)\right].
    \label{impulse}
\end{equation}
We have opted to use distance rather than time to parameterize motion, $T\equiv s_p(T)/c$, in order to simplify certain derivatives and expressions. The first two terms are the impulse associated with added mass while the last term is the impulse due to the shed vortex. The force on the plate with respect to the ground frame is,
\begin{equation}
    F=-\rho\frac{dI}{dt}.
    \label{force}
\end{equation}
We define the force coefficient in the reference frame of the plate as $c_{F_{plate}}\equiv 2Fe^{-i(\beta_p(T)+\alpha_p(T))}/(\rho c \dot{s}_p(T)^2)$, so that the real part is the force along the plate and the imaginary part is the force normal to the plate. Using this definition as well as equations \ref{force} and \ref{impulse} and our definitions for dimensionless quantities,
{\scriptsize
\begin{equation}
    c_{F_{plate}}(T) = \frac{i}{2\mathbb{U}_p(T)}e^{-i(\beta_p(T)+\alpha_p(T))}\frac{d}{dT}\left[e^{i(\beta_p(T)+\alpha_p(T))}\left[\pi\mathbb{U}_p(T)\sin(\alpha_p(T))-\pi \mathbb{U}_p(T)(p-1/2)\kappa_p(T)-\Gamma\left(\zeta_\Gamma(T)-\frac{1}{\overline{\zeta}_\Gamma(T)}\right)\right]\right].
    \label{drag coefficient 2 vortex}
\end{equation}
}
If the vortex has been advected extremely far from the plate it will not move with respect to the ground frame. Therefore, changes in $\zeta_\Gamma(T)$ will be purely a result of motion of the plate with respect to the ground frame as the TEV is fixed in the ground frame. Furthermore $1/\overline{\zeta}_\Gamma(T) \to 0$ in this limit so the image vortex's motion will not contribute to changes in impulse. The motion of the plate can be converted into the $z$-plane from the $\zeta$-plane with a reduced expression since the vortex is far away from the plate: $\zeta_\Gamma(T)\approx 4\left(\frac{z_\Gamma(T)}{c}-(p-1/2)\right)$. Meanwhile, the location of the center of the plate in the ground frame is $\tilde{z}_c(T)=\tilde{z}_p(T)+c(p-1/2)e^{i(\beta_p(T)+\alpha_p(T))}$ where $\tilde{z}_p(T)$ is the location of the point of rotation with respect to the ground frame. Also, the conversion between $\tilde{z}$ and $z$ reference frames is $z_\Gamma(T)=(\tilde{z}_\Gamma-\tilde{z}_c)e^{-i(\beta_p(T)+\alpha_p(T))}$. Therefore the expression of $\zeta_\Gamma(T)$ in terms of the location of the plate is,
\begin{equation}
    \zeta_\Gamma(T)\approx\frac{4}{c}(\tilde{z}_\Gamma-\tilde{z}_p(T))e^{-i(\beta_p(T)+\alpha_p(T))}-4(p-1/2).
\end{equation}
The general expression for the force coefficient in the plate-fixed frame is therefore,
\begin{multline}
    c_{F_{plate}}(T) = \frac{i}{2\mathbb{U}_p(T)}e^{-i(\beta_p(T)+\alpha_p(T))}\frac{d}{dT}\left[e^{i(\beta_p(T)+\alpha_p(T))}\left[\pi\mathbb{U}_p(T)\sin(\alpha_p(T))\right.\right.\\
    \left.\left.-\pi \mathbb{U}_p(T)(p-1/2)\kappa_p(T)-\Gamma\left(\frac{4}{c}(\tilde{z}_\Gamma(T)-\tilde{z}_p(T))e^{-i(\beta_p(T)+\alpha_p(T))}-4(p-1/2)\right)\right]\right].
    \label{drag coefficient general}
\end{multline}
To correctly use this equation it is necessary to be able to take the time derivatives of all relevant terms. By definition the motion of the plate's point of rotation is $\frac{d\tilde{z}_p(T)}{dt}=\dot{s}_p(T)e^{i\beta_p(T)}$, so $\frac{d\tilde{z}_p(T)}{dT}=ce^{i\beta_p(T)}$. Meanwhile, $\frac{d(\beta_p(T)+\alpha_p(T))}{dT}=\kappa_p(T)$ and $\frac{d\alpha_p(T)}{dT}=\kappa_{\alpha_p}(T)$. The derivative of the other terms are self-evident. Evaluating the derivatives results in, 
\begin{multline}
    c_F(T) = \frac{i\pi}{2}\left[\frac{\mathbb{U}'_p(T)}{\mathbb{U}_p(T)}\sin(\alpha_p(T))+\cos(\alpha_p(T))\kappa_{\alpha_p}(T)-i\sin(\alpha_p(T))\kappa_p(T)\right]e^{-i\alpha_p(T)}\\
    -\frac{i\pi(p-1/2)}{2}\left(\frac{\mathbb{U}'_p(T)}{\mathbb{U}_p(T)}\kappa_p(T)+\kappa'_p(T)-i\kappa_p^2(T)\right)e^{-i\alpha_p(T)}+2(p-1/2)\frac{\Gamma\kappa_p(T)}{\mathbb{U}_p(T)}e^{-i\alpha_p(T)}+\frac{2\Gamma i}{\mathbb{U}_p(T)}
    \label{drag coefficient general 2}
\end{multline}
This expression has been converted from the body-fixed frame into a motion-aligned frame using definition $c_F\equiv -e^{-i\alpha_p(T)}\overline{c}_{F_p}$. This frame is a complex representation of the standard lift and drag coefficients: $c_F= c_D+ic_L$.

\section{Asymptotic Equivalence of Lift Formulas}
\label{Asymptotic Equivalence of Lift Formulas}
Here we show that the lift component of \cref{coefficient of force suppression general} is equivalent to quasi-steady lift formula 6 and Theodorsen lift formula 7 in \citet{mcgowan2011investigations} (with the steady state of equation 5 added in), in the limit of small and slow pitch and heave. If we decompose the airfoil's velocity vector into $u_\infty$ and $\dot{h}(T)$ components we find \cref{Kutta-Joukowski force} becomes,
\begin{equation}
    c_{F_J}=2\pi\left(\sin(\alpha_p)+\left(\frac{3}{4}-p\right)\frac{c\Omega(T)}{\sqrt{u_\infty^2+\dot{h}^2}}\right)i.
\end{equation}
When $\dot{h}/u_\infty\ll 1$ and $\alpha_p\ll 1$ the first order Maclaurin series becomes,
\begin{equation}
    c_{F_J}\approx 2\pi\left(\alpha_p+\left(\frac{3}{4}-p\right)\frac{c\Omega(T)}{u_\infty}\right)i.
\end{equation}
The notation conversion to \citet{mcgowan2011investigations} is $\alpha_p+\beta_p=\alpha_m+\theta$ so that $\Omega=\dot{\theta}$ and $\alpha_p=\alpha_m+\theta-\arctan{(\frac{\dot{h}}{u_\infty})}$. Hence the Maclaurin expansion of the expression using pitch-heave coordinates is,
\begin{equation}
    c_{F_J}\approx 2\pi\left(\alpha_m+\theta-\frac{\dot{h}}{u_\infty}+\left(\frac{3}{4}-p\right)\frac{c\dot{\theta}}{u_\infty}\right)i,
    \label{Joukowski linear}
\end{equation}
which is identical to the sum of their equation 5 and 6, except they have a single, constant, lift offset term as well due to their airfoil having non-zero camber (they also normalize force by $u_\infty$ rather than $\dot{s}_p(T)$, but these are the same to first order). The lift component of the added mass portion of \cref{coefficient of force suppression general} is,
\begin{multline}
    \text{Im}[c_{F_{Add.}}]=\frac{\pi}{2}\left[\sin(\alpha_p)\cos(\alpha_p)\frac{c\dot{h}\ddot{h}}{(u_\infty^2+\dot{h}^2)^{3/2}}+\cos^2(\alpha_p)\frac{c\dot{\alpha}_p}{\sqrt{u_\infty^2+\dot{h}^2}}-\sin^2(\alpha_p)\frac{c\Omega(T)}{\sqrt{u_\infty^2+\dot{h}^2}}\right]\\
    -\frac{\pi}{2}(p-1/2)\left[\cos(\alpha_p)\frac{c^2\dot{\theta}\dot{h}\ddot{h}}{\left(u_\infty^2+\dot{h}^2\right)^2}+c^2\cos(\alpha_p)\frac{\ddot{\theta}\sqrt{u_\infty^2+\dot{h}^2}-\frac{\dot{\theta}\dot{h}\ddot{h}}{\sqrt{u_\infty^2+\dot{h}^2}}}{\left(u_\infty^2+\dot{h}^2\right)^{3/2}}-\sin(\alpha_p)\frac{c^2\dot{\theta}^2}{u_\infty^2+\dot{h}^2}\right]\\
    -2\pi(p-1/2)\sin(\alpha_p)\left[\sin(\alpha_p)-(p-3/4)\frac{c\dot{\theta}}{\sqrt{u_\infty^2+\dot{h}^2}}\right]\frac{c\dot{\theta}}{\sqrt{u_\infty^2+\dot{h}^2}}
\end{multline}
The first and last terms of the first bracketed group have leading order greater than 1 in the expansion so they drop out leaving only the middle term. Only the $\ddot{\theta}$ term survives in the second bracketed group, and no terms survive in the last bracketed group. The resulting expansion to first order is then,
\begin{equation}
    \text{Im}[c_{F_{Add.}}]\approx \frac{\pi c}{2}\frac{\dot{\alpha}_p}{u_\infty}-\frac{\pi}{2}(p-1/2)c^2\frac{\ddot{\theta}}{u_\infty^2},
\end{equation}
and this becomes,
\begin{equation}
    \text{Im}[c_{F_{Add.}}]\approx \frac{\pi c}{2}\left[-\frac{\ddot{h}}{u_\infty^2}+\frac{\dot{\theta}}{u_\infty}-(p-1/2)c\frac{\ddot{\theta}}{u_\infty^2}\right].
    \label{added mass linear}
\end{equation}
The sum of equations 5 and 7 in \citet{mcgowan2011investigations} is precisely the sum of \cref{Joukowski linear} and \cref{added mass linear} when the Theodorsen function, $C(k)$, is exactly 1.
\end{document}